\documentclass[iop]{emulateapj}

\usepackage{natbib}
\usepackage{amsmath, amssymb}
\usepackage{times}
\usepackage{apjfonts, graphicx, graphics}
\usepackage[breaklinks,colorlinks,urlcolor=blue,citecolor=blue]{hyperref}
\usepackage[all]{hypcap}
\usepackage{aas_macros}
\usepackage{subfigure}
\usepackage{url}

\shorttitle{Cooling+Heating Flows in Galaxy clusters}
\shortauthors{}

\begin{document}
\title{Cooling+Heating Flows in Galaxy clusters: \\ Turbulent heating, spectral modelling, and cooling efficiency}
\author{Mohammad H. Zhoolideh Haghighi$^{1,2,3}$}
\author{Niayesh Afshordi$^{2,3,4}$}
\author{Habib. G. Khosroshahi$^{1}$}

\affil{
	$^{1}$School of Astronomy, Institute for Research in Fundamental Sciences (IPM), Tehran, 19395-5746, Iran \\
	$^{2}$Perimeter Institute for Theoretical Physics, 31 Carolines St. North, Waterloo, ON, N2L 2Y5, Canada \\
	$^{3}$Department of Physics, Sharif University of Technology, P.O. Box 11155-9161, Tehran, Iran \\
	$^{4}$Department of Physics and Astronomy, University of Waterloo, 200 University Ave. West, Waterloo, ON, N2L 3G1, Canada \\	
}

\begin{abstract}
	
	The discrepancy between expected and observed cooling rates of X-ray emitting gas has led to the {\it cooling flow problem} at the cores of clusters of galaxies. A variety of models have been proposed to model the observed X-ray spectra and resolve the cooling flow problem, which involves heating the cold gas through different mechanisms. As a result, realistic models of X-ray spectra of galaxy clusters need to involve both heating {\it and} cooling mechanisms. In this paper, we argue that the heating time-scale is set by the magnetohydrodynamic (MHD) turbulent viscous heating for the Intracluster plasma, parametrised by the Shakura-Sunyaev viscosity parameter, $\alpha$. Using a cooling+heating flow model, we show that a value of $\alpha\simeq 0.05$ (with 10\% scatter) provides improved fits to the X-ray spectra of cooling flow, while at the same time, predicting reasonable cooling efficiency, $\epsilon_{cool} = 0.33^{+0.63}_{-0.15}$. Our inferred values for $\alpha$ based on X-ray spectra are also in line with direct measurements of turbulent pressure in simulations and observations of galaxy clusters. This simple picture unifies astrophysical accretion, as a balance of MHD turbulent heating and cooling, across more than 16 orders of magnitudes in scale, from neutron stars to galaxy clusters.  
	
\end{abstract}

\keywords{
	intracluster medium 
	galaxies: clusters: individual (Hydra~A, A2029, A2199, A496, A85) 
	galaxies: cooling flow problem: active galactic nucleus (AGN):  
}

\altaffiltext{*}{
	\href{mailto:mzhoolideh@ipm.ir}{mzhoolideh@ipm.ir}
}

\section{Introduction}
The Intracluster Medium (ICM) of galaxy clusters consists of a plasma that is almost entirely ionized. This hot plasma radiates mostly in X-ray band which leads to a significant cooling of the ICM. At constant pressure, the cooling time of a plasma is the gas enthalpy divided by the energy lost per unit volume per unit time: 
\begin{equation}
t_{\rm cool}\equiv \frac{5n k_{\rm B}T}{2n_e n_H \Lambda(T,Z)},\label{tcool}
\end{equation}
 where $\Lambda(T, Z)$ is the cooling function in terms of temperature $T$ and metallicity $Z$, $n$ is the particle number density, and $k_{\rm B}$ is the Boltzmann's constant. In the cores of clusters, the cooling time
dips below $5\times10^8$ yr, i.e.
the inferred radiative cooling time of the gas in the central part, where X-ray emission is sharply peaked, is much shorter than the age of the cluster, which suggests the existence of cooling flow. The standard cooling flow model can be derived by combining continuity, Navier-Stokes and energy conservation equation that, after simplification, leads to:
\begin{equation}
\frac{dL_{X}}{dT} = \dot{M}\left(\frac{5k_{\rm B}}{2\mu m_p}-\frac{1 dp}{\rho dT}\right).
\end{equation}
In the case of constant pressure, we get the standard isobaric cooling flow model:
\begin{equation}
\frac{dL_{X}}{dT_{K}} = \frac{5\dot{M}k_{\rm B}}{2\mu m_{p}}.
\label{dL/dT}
\end{equation}

X-ray spectroscopy has demonstrated that this model is inadequate
and additional heating or cooling mechanisms should be incorporated into the model.
Moreover, X-ray spectroscopy shows that the temperature drop toward the center is limited to about a factor of three. The cooling seems to be frozen precisely in the region where we expect more rapid cooling. In general, it appears that there is no strong evidence for any significant amount of cold X-ray emitting gas (below 1/3 of the maximum temperature) in any cluster \cite[][]{Peterson06}.
 
 There exist different manifestations of the {\it cooling-flow problem}: According to \citet{Peterson03}, there is the soft X-ray cooling-flow problem and the mass sink cooling-flow problem. The soft X-ray cooling-flow problem refers to the discrepancy seen between the predicted and observed soft X-ray spectrum, e.g. the lack of expected emission lines from a gas cooling to low temperatures at the core of the cluster. The mass sink cooling-flow problem refers to the lack of colossal mass deposition in cooling clusters from the hypothesized cooling-flow plasma.
 
 Many mechanisms have been proposed to prevent the gas from cooling to low temperatures at the centers of cooling flows,  {such as the electron thermal conduction \citep{Zakamska2003} , Mechanical heating of infalling gas in dense core systems \citep{Khosroshahi2004} and turbulent heating \citep{2014Natur.515...85Z}, though the lead suspect amongst them is mechanical heating by Active Galactic Nuclei (AGN).} AGN outbursts produce winds and intense radiation that can heat the gas. Produced weak shocks  delay cooling of gas by reducing gas density and increasing the total energy  \cite[][]{David2001,McNamara05,Forman05} or by compensating lost entropy of the gas \cite[][]{Fabian05}. Moreover,  viscous damping of sound waves generated by repeated AGN outbursts may represent a significant source of heating \cite[][]{Fabian03}. 
Direct evidence for these sound waves came from the spectra observed by the Hitomi X-ray satellite, which measured the plasma's line-of-sight velocity dispersion of $164\pm10 $ km/s within the core of Perseus cluster. This supports the hypothesis that turbulent dissipation of kinetic energy can supply enough heat to offset gas from cooling \citep{Hitomi2016}. 

In this paper,we provide a simple yet accurate thermodynamic model for cooling{\it +heating} (or CpH) flows, which captures the balance between turbulent heating and cooling in cluster cores \citep[e.g.,][]{2014Natur.515...85Z}. We then show that the model can simultaneously explain the X-ray spectra and the observed turbulent energy of the cluster cores, using a single parameter $\alpha \simeq 0.1$, for the Shakura-Sunyaev viscosity parameter, while at the same time, predict reasonable cooling efficiency. As such, this picture also unifies astrophysical accretion across 16 orders of magnitude in scale, from kilometres (around neutron stars and stellar black holes) to kiloparsecs (in cores of galaxy clusters).

\section{Data and spectral Extraction}\label{Section:Sample}

For this study, we use a sample of galaxy clusters presented by \cite{Hogan17}. The sample consists of 5 galaxy clusters observed with Chandra X-ray Observatory over long exposure times. All five clusters have a central cooling time $\leq 1\times 10^9$ yrs \citep{Cavagnolo09} suitable for our intended analysis. The data are obtained from the Chandra imaging  online repository and analyzed using CIAO version 4.7. Bad pixels are masked out using the bad pixel
map provided by the pipeline. Background flares are removed, and point sources are identified with the CIAO task WAVDETECT and masked out in all subsequent analysis. Finally, the blank-sky backgrounds are extracted 
for each target, and the images are prepared in the energy range 0.5--7.0~keV. In addition, cavities and filaments within ICM were masked clear, since these regions are usually out of equilibrium.

Because the cooling instabilities usually  occur at small ($\lesssim$10~kpc) radii, we desire finely binned spectra in the central cluster regions. Our example clusters have deep Chandra data; as a result, choosing of annuli for spectral extraction is limited by resolution rather than the number of counts.

For each example cluster, concentric circular annuli are centered at the cluster center. The width of the central annulus is 3 pixels, where each pixel is 0.492~arcsec across. The width of each annulus increases in turn by 1-pixel until the sixth annulus, beyond which the width of each annulus is 1.5 times the width of the previous one with a total number of 10 annuli per source.  
For each \textsc{obsid} we have spectra alongside response matrix files (RMFs) and auxiliary response files (ARFs). We keep  spectra separate before fitting them, but at the time of running the \textsc{xspec} we load them simultaneously.

Since emission from outer parts of the ICM affects and contaminates inter parts of spectra, we use deprojected spectra to obtain more accurate data.
In order to fit observed data we load the extracted spectra for each cluster with their matched response files into \textsc{xspec} version 12.9.1 \citep{Arnaud96} and use fixed values of $N_H$ reported in \citep{Main15}.

\begin{table*}
	\centering
		
	\begin{tabular}{|c||c||c||c||c||c||c||c|}
		\hline\hline
		Cluster & $N_{H}$ &  z    & Scale    &  Observation IDs                  &   Exposure (ks)  & $\dot{M}_{SFR}$  &   $\dot{M}_{cool}$   \\
		        &   ($10^{22}cm^{-2})$   &  & (kpc/'') &                                   &   Cleaned        & $M_{\odot}yr^{-1}$ &      $M_{\odot}yr^{-1}$        \\
		\hline                                                                                                                                             
		A2029  &0.033 & 0.0773 & 1.464    & 891, 4977, 6101                   &  103.31      & $0.8^{0.1}_{0.09}$   &       269.2   $ \pm$ 1.1      \\
		A2199  & 0.039 & 0.0302 & 0.605    & 10748, 10803, 10804, 10805        & 119.61      & $1^{9}_{1}$    &       47.9  $ \pm$ 1.1     \\
		A496  & 0.040  & 0.0329 & 0.656    & 931, 4976                         &  62.75      & $0.18^{0.01}_{0.01}$       &    56.2 $ \pm$ 1.1         \\
		A85   & 0.039  & 0.0551 & 1.071    & 904, 15173, 15174, 16263, 16264   & 193.64        & $0.1^{2.5}_{0.1}$     &   87.1 $ \pm$ 1.0           \\
		Hydra~A & 0.043 & 0.0550 & 1.069    & 4969, 4970                        &  163.79    & $4^{7}_{2}$        &      109.6  $ \pm$ 1.0       \\
		\hline
	\end{tabular}
\caption{ {Sample clusters from Chandra data \citep{Hogan17}. Standard cosmology with H0= 70 $km ~ s^{-1} ~ Mpc^{-1}$ has been used and scales are angular. The observed star formation rate ,$\dot{M}_{SFR}$, and classical cooling rate, $\dot{M}_{cool}$,  is obtained from \cite{McDonald2018} and the K-band magnitude are collected from \cite{ngc6166} and \cite{2mass} }.
	}
	\label{Observations_Table}
\end{table*}

\section{Modified cooling+heating (CpH) flow model: cooling v.s. sound crossing}

AGNs outburst and jets can pump energy into the ICM \citep{Nulsen2007}. This can be done by shock waves or sound wave deposition close to the AGN.  We introduce a new timescale, $t_{\rm heat}$, which represents time scale of the energy injection into the system by viscous heating. To estimate the heating time, we note that waves (and weak shocks) produced by AGcluster ofNs can travel at most by speed of sound. As a result, sound crossing time is the shortest time scale in the ICM. We further assume that heating or viscous time should be a multiple of sound crossing time 
\begin{equation}
t_{\rm heat} =  \alpha^{-3/2} t_{\rm sound} = \frac{R}{\alpha^{3/2} c_s} = R \sqrt{3 \mu m_p \over 5\alpha^3 k_{\rm B} T} , 
\end{equation}
where $R$ is the distance to the center of the cluster, and  $\alpha <1$. The $\alpha$ parameter quantifies the ratio of turbulent/magnetic to thermal energy, and is 
very similar to the Shakura-Sunyaev viscosity parameter in accretion disks \citep{1973A&A....24..337S}, as in a turbulent medium equipartition implies:
\begin{equation}
\langle v^2 \rangle \sim \langle v^2_A \rangle \sim \alpha c^2_s,
\end{equation}
where $v$ and $v_A$ are turbulent and Alfven speeds. The heating time is then given by the ratio of thermal energy $n k_{\rm B} T$ by turbulent heating rate $ \rho \langle v^2 \rangle \frac{\langle v^2 \rangle^{1/2}}{R}$:
\begin{equation}
t_{\rm heat} \sim \frac{n k_{\rm B} T} {\rho \langle v^2 \rangle^{3/2}/R } \sim \frac{R c^2_s}{\langle v^2 \rangle^{3/2}} \sim \frac{R}{\alpha^{3/2} c_s} =  \alpha^{-3/2} t_{\rm sound}. \label{theat}
\end{equation}
The viscosity parameter typically takes a value of $\alpha \sim 0.01-0.1$ in magnetohydrodynamic (MHD) simulations of weakly magnetized plasmas \citep[e.g.,][]{2016MNRAS.457..857S}.  

\begin{figure}
	\begin{center}
		\begin{tabular}{cc}
			\includegraphics[scale = 0.25 ]{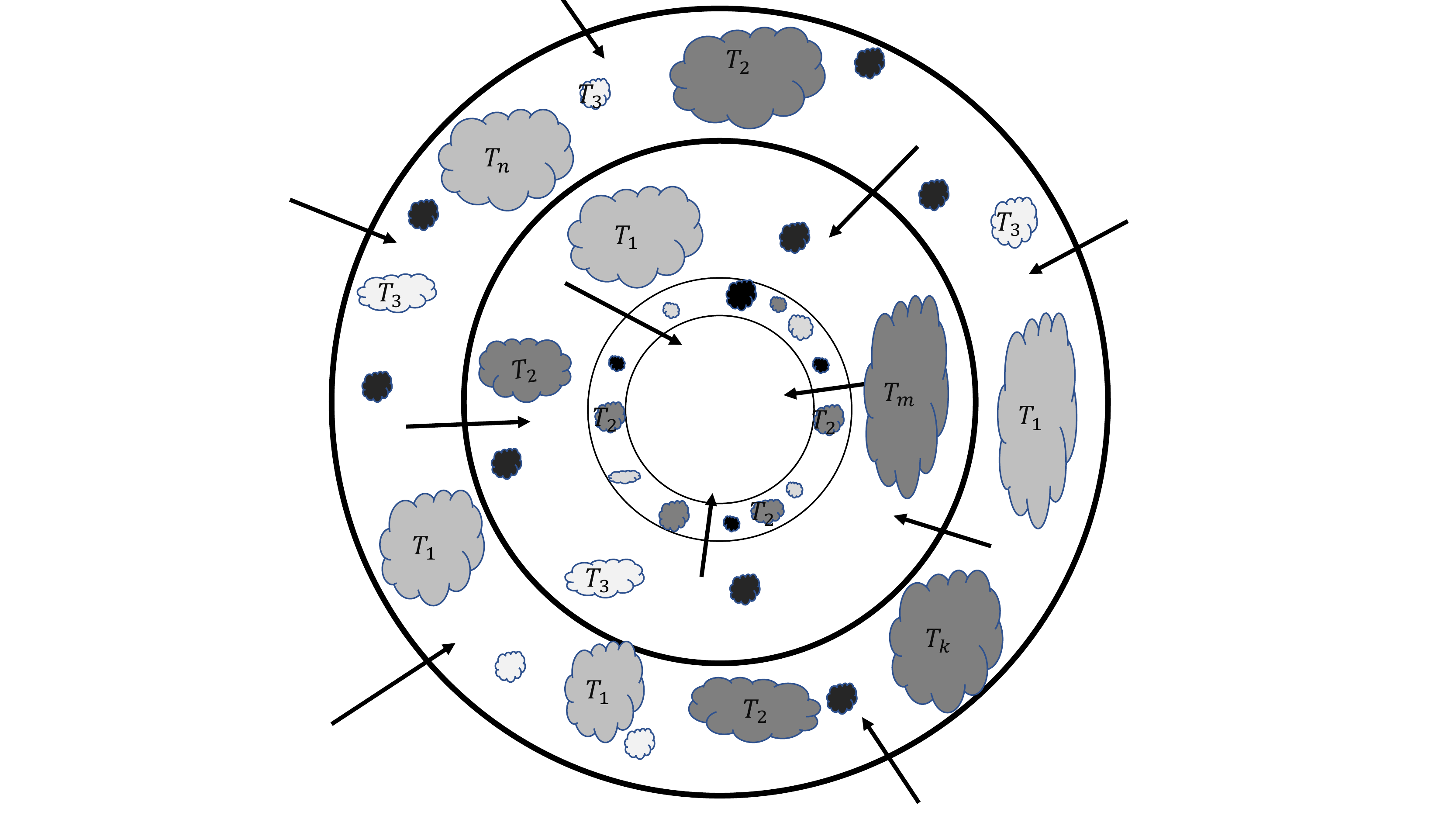}\\
		\end{tabular}
	\end{center}
	\caption{Schematic illustration of cooling flow in each annulus. Here, we model temperature distribution within each annulus as a result of in-situ cooling and heating processes.}
	\label{coldflow}
\end{figure}

Now, the central idea of our proposal is that the main driver of thermal distribution in ICM is a balance of cooling and heating, governed by $t_{\rm cool}/t_{\rm sound}$ \citep[in contrast to, e.g., thermal instability determined by  $t_{\rm cool}/t_{\rm free-fall}$][]{McCourt12,Hogan17}. Since $t_{\rm cool}$ drops faster than $T^{3/2}$ (at constant pressure; see below) while  $t_{\rm heat} \propto t_{\rm sound}$ grows as $T^{-1/2}$ at low temperatures, the cold gas cools copiously, as in the standard cooling flows. However, the heating would win over the cooling for high temperatures. As a result, there is thermodynamic equilibrium at the temperature $T^*$ where $t_{\rm cool}(T^*) \sim t_{\rm heat}(T^*)$. In a (nearly) steady state (e.g., close to the cluster core), most of the gas would sit near this temperature \footnote{Even though this is an unstable equilibrium, random turbulent motion can make it semi-stable, similar to the inverted pendulum with an oscillating tip \citep[Kapitza's pendulum;][]{1969mech.book.....L}.}. However, gas with $T \ll T^*$ would cool down to form atomic or molecular gas, and eventually stars. On the other hand gas with $T \gg T^*$ would heat up and eventually feed the cosmic ray population, through Fermi acceleration.  This in-situ mass loss within each annulus should be replenished by a slow accretion/inflow of plasma from cluster outskirts, in steady state (see Figure \ref{coldflow} for a visual representation).

To see this more quantitatively, let's begin with the cooling flow model. In the case of standard isobaric cooling flow model, we have:

\begin{equation}
\frac{5k_{\rm B} n}{2}\frac{dT}{dt}=-n_e n_H\Lambda(T,Z)
\label{1}
\end{equation}
or 
\begin{equation}
\frac{d\ln T}{dt}=-\frac{1}{t_{\rm cool}}.
\label{2}
\end{equation}

To modify this, we add a heating term to the equation (\ref{2}) and rewrite it as:
\begin{equation}
\frac{d\ln T}{dt}=-\frac{1}{t_{\rm cool}} + \frac{1}{t_{\rm heat}} 
\end{equation}

 The mass-weighted probability distribution of $\ln T$ should satisfy the conservation equation:
\begin{equation}
\frac{\partial P(\ln T)}{\partial t}+ \frac{\partial}{\partial \ln T} \left[P(\ln T)\frac{d\ln T}{d t}\right]=0,
\end{equation}
which in steady state yields: 
\begin{equation}
P(\ln T)\frac{d\ln T}{dt}= A = {\rm const.}
\end{equation}

Therefore the probability distribution takes the following form:

\begin{equation}
P_{CpH}(\ln T)= A \frac{t_{\rm cool}}{|1-\frac{t_{\rm cool}}{t_{\rm heat}}|},
\label{prob-dist}
\end{equation}
where $A$ is a normalization constant that is fixed by requiring total integrated probability \footnote{We should note that, even though this integral is formally divergent at $t_{\rm cool}=  t_{\rm heat}$, the divergence is only logarithmic, and is presumably regularized by stochastic turbulent motion. For our calculation, the integral is regularized by the finite temperature bins in the spectral modeling. However, due to the logarithmic nature of divergence, the choice of binning has little effect on our results.} is 1: 
\begin{equation}
A^{-1} = \int d\ln T \frac{t_{\rm cool}}{|1-\frac{t_{\rm cool}}{t_{\rm heat}}|}.
\end{equation}
Now, as discussed above, we see from Equation (\ref{prob-dist}) that when $t_{\rm cool} \ll t_{\rm heat}$ (or $T \ll T^*$), we obtain the standard cooling flow model. In contrast, if $t_{\rm cool} \gg t_{\rm heat}$ (or $T \gg T^*$) we have $P(\ln T) \approx t_{\rm heat}$.

\begin{figure}
	\begin{center}
		\begin{tabular}{cc}
			\includegraphics[width=70mm]{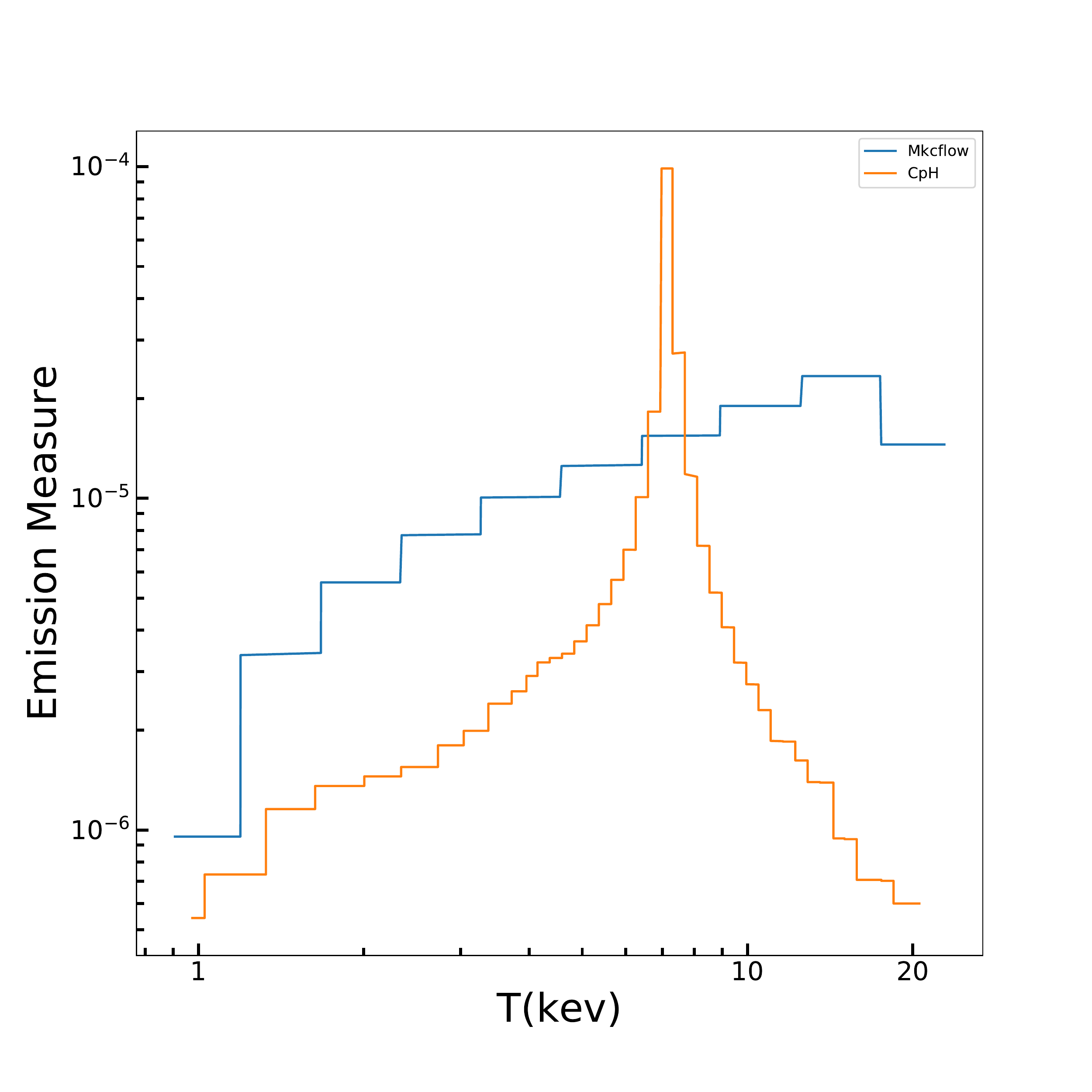}\\
		\end{tabular}
	\end{center}
\caption{ The emission measure for seventh annulus of A2029 with using {\it mkcflow} model and our CpH model.}
\label{emission measure}
\end{figure}

\section{A New Spectral Model}

In the standard cooling flow model, the spectrum in steady state can be calculated using:
\begin{equation}
dL_{\rm cool}=n_{e}n_{H}\Lambda(T,Z)dV=\frac{5\dot{M}k_{\rm B}}{2\mu m_{p}}dT,
\end{equation}
where $\dot{M} $ is the mass deposition rate, and $\mu$ is the mean molecular weight. We can re-express this equation in terms of a differential emission measure $dEM = n_{e}n_{H}dV$ which captures the differential distribution of plasma across temperatures: 
\begin{equation}
\frac{dEM}{dT} = \frac{5\dot{M}k_{\rm B}}{2\mu m_{p}\Lambda(T,Z)}.
\end{equation}
We can now convolve this distribution with the energy dependent line power $\frac{d\kappa}{dE}(E,T,Z)$ to
produce an X-ray spectrum, which can be compared with observations. The spectral source model as a function of emission measure will be:
\begin{equation}
\frac{d\epsilon}{dE} = \int_{0}^{T_{\rm max}}\frac{dEM}{dT}\frac{d\kappa}{dE}(E,T,Z)dT,
\label{cooling-Spec}
\end{equation} 
where 
\begin{equation}
\Lambda(T,Z)= \int_0^\infty dE \frac{d\kappa}{dE}(E,T,Z) E.
\end{equation}

Equation (\ref{cooling-Spec}), which is the prediction of the standard cooling flow model, cannot provide a good fit to the X-ray spectra of galaxy cluster cores (or outskirts) \cite[][]{Peterson06}. In practice, it is common to use a single (or multi-) temperature model (so-called ``{\it mekal}'' in \textsc{xspec}) to fit the X-ray spectra, even though this cannot be physically justified given the short cooling times in cluster cores.

Our proposal to solve the problem is to apply the probability distribution \ref{prob-dist} introduced in the previous section, which captures both heating and cooling in the flow. By plugging in the explicit expressions for of $t_{\rm cool}$ (Eq. \ref{tcool}), $t_{\rm heat}$ (Eq. \ref{theat}) and pressure $p =nkT$ we find:
\begin{equation}
P_{CpH}(\ln T)= A \frac{t_{\rm cool}}{|1-\frac{f(T)}{f(T^*)}|},
\label{norm}
\end{equation}
where 
\begin{equation}
f(T) \equiv \frac{T^{5/2}}{\Lambda(T,Z)},~ {\rm and}~~~
f(T^*)=\frac{n_e n_H}{n^2}\sqrt{12 \mu m_p p^2 R^2 \over 125 \alpha^3 k_{\rm B}^5}
\label{fTstar}
\end{equation}

We can now modify the cooling flow emission measure $\frac{dEM}{dT} \rightarrow  \frac{1}{|1-\frac{f(T)}{f(T^*)}|} \frac{ dEM}{dT}$ and apply it in the spectral model (\ref{cooling-Spec}):
\begin{equation}
\frac{d\epsilon}{dE} = \int_{0}^{T_{max}}\frac{1}{|1-\frac{f(T)}{f(T^*)}|} \frac{dEM}{dT}\frac{d\kappa}{dE}(E,T,Z)dT.
\label{our-Spec}
\end{equation}

We fit the observed spectra after implementing our CpH model into the \textsc{xspec}. To do this, we modify the emission measure of the standard cooling flow (or ``{\it mkcflow}'') model. \footnote{We modify the source codes of {\it mkcflow} model in \textsc{xspec} to match our desired emission measure.} To follow the changes in the emission measure after we implement our model into the \textsc{xspec}, we present an example of our best-fit emission measure of the cooling flow model, compared to our model, for the same cluster and the same annulus in Figure \ref{emission measure}. In contrast to the {\it mkcflow} mode, which has a smooth emission measure, our model has a clear peak in temperature (where $T=T^*$ or $t_{\rm heat} = t_{\rm cool}$), as well as extended tails. 

In order to fit observed Chandra spectra, $T^*$ is treated as a free parameter in \textsc{xspec}, while we fixed the lower and the upper limits of the integration $T_{\rm min} = 10^{-2}$ keV and $T_{\rm max} = 50$ keV. As such, our spectral model has the same number of parameters as a single-temperature (or {\it mekal}) model. It is worth mentioning that if the lower limit value of {\it mkcflow} model is set to such a small value, it is impossible to fit the observed data. In contrast, as we see below, we can find a good fit to data using the modified emission measure (\ref{our-Spec}). 

To obtain $\alpha$ we first compute the pressure , $p$ within each annulus of volume $V = 4/3 \pi (r^3_{out}-r^3_{in})$, as follows:

	\begin{equation}
	P(\ln T) = \frac{dM}{M_{tot} d\ln T}
	\label{p_dm}
	\end{equation}
	
	\begin{equation}
	\frac{dM}{d\ln T}= \mu m_p n\frac{dV}{d\ln T} = \mu m_p \frac{p}{k_B T}\frac{dV}{d\ln T}
	\label{dm_dlnT}
	\end{equation}
By eliminating $dM$ in Eq. (\ref{dm_dlnT}) using Eq. (\ref{p_dm})and integrating over volume we get the following relation for $M_{tot}$:

\begin{equation}
{M}_{tot} = \frac{\mu m_p p V}{\int k_B T P(\ln T) d\ln T}
\end{equation}
We notice that normalization of the Eq. (\ref{norm}), $A$, has the inverse time dimension and as a result we define our  mass deposition rate in each annulus as $\dot{M}_{cool} = AM_{tot}$. As a result after some simplification we can express pressure in terms of $\dot{M}_{cool}$  and $T^*$:
\begin{equation}
p^2 = \frac{5 \dot{M}}{2 \mu m_p V} \int_{0}^{T_{max}}\frac{k_B^3 T^3 d\ln T}{\Lambda (T,Z)|1-\frac{f(T)}{f(T^*)}|} 
\end{equation}

We can further express $n_{e}$ and $n_{H}$ in terms of the total number density $n$, using  $n = n_e + n_H + n_{He}$,  $ X = n_H/(n_H+4n_{He})$ and $n_e = n_H + 2n_{He}$, which yield
\begin{equation}
n_{e} = \frac{2X+2}{5X+3}n, \hspace{1.5cm} n_{H} = \frac{4X}{5X+3}n, \label{n_X}
\end{equation}
where $X$ is the hydrogen mass fraction and we shall consider $X \approx 0.75$.
Now, by eliminating pressure $p$ in the following equation for $\alpha$  (\ref{ALPHA}) and $f(T^*)$ (Eq. \ref{fTstar}), using Eq. (\ref{n_X}) for $n_e$ and $n_H$, we find
$\alpha$ which gives the value of viscosity parameter, within the annulus at radius $R$ and volume $V$, in terms of the X-ray observables $T^*$ and $\dot{M}$.

\begin{equation}
\alpha = \Big[\frac{125  k^5_{\rm B} n^4 f(T^*)^2}{12 n_e^2 n_H^2 \mu m_p p^2 R^2}\Big]^{-1/3} 
\label{ALPHA}
\end{equation}

 We name our model Cooling plus Heating (CpH) and it is publicly available at: 
	{\url{https://heasarc.gsfc.nasa.gov/xanadu/xspec/manual/node158.html}}

\section{Results}

In this section, we probe our CpH model using the Chandra clusters sample and demonstrate that  they are superior (or comparable) fits in cluster cores, in comparison to single-temperature {\it mekal} models. We then briefly discuss the implications for the cooling efficiency and turbulent viscosity in ICM.

We use \textsc{xspec} to fit Chandra X-ray data. For the single temperature model (phabs$\times${\it mekal}) and our CpH model (phabs$\times$ CpH). We fix abundance to the solar and also fix the hydrogen column densities $N_{H}$ and redshifts to the values provided in Table (\ref{Observations_Table}). We ran Markov Chain Monte Carlo (MCMC) to find best-fit parameters. The best-fit parameter for single temperature model is $T$ and for our CpH model is  $T^*$ (indicating the peak of the CpH probability distribution). The goodness of the fit and the best-fit parameters of our model, as well as the single temperature model, are provided in Table \ref{Best_fit_tabel}.
We notice that the peaks of the temperature distribution in our best-fit models, $T^*$, happen to be close to the best-fit $T$ in the single-temperature models. However, typically our model provides a better (or comparable) fit to data in cluster cores (with an acceptable $\chi^2$ for the number of data points). In the cluster outskirts, where the assumption of a steady state cooling/heating flow is not valid, none of the models provide a good fit to the data.  While the CpH and the {\it mekal} models provide satisfactory fits to X-ray data for the same annuli, the CpH model is preferred at $\Delta \chi^2= -56.9$ (or $7.5 \sigma$ level) if we combine all the annuli with satisfactory fits.

\begin{figure}
	\includegraphics[scale = 0.35]{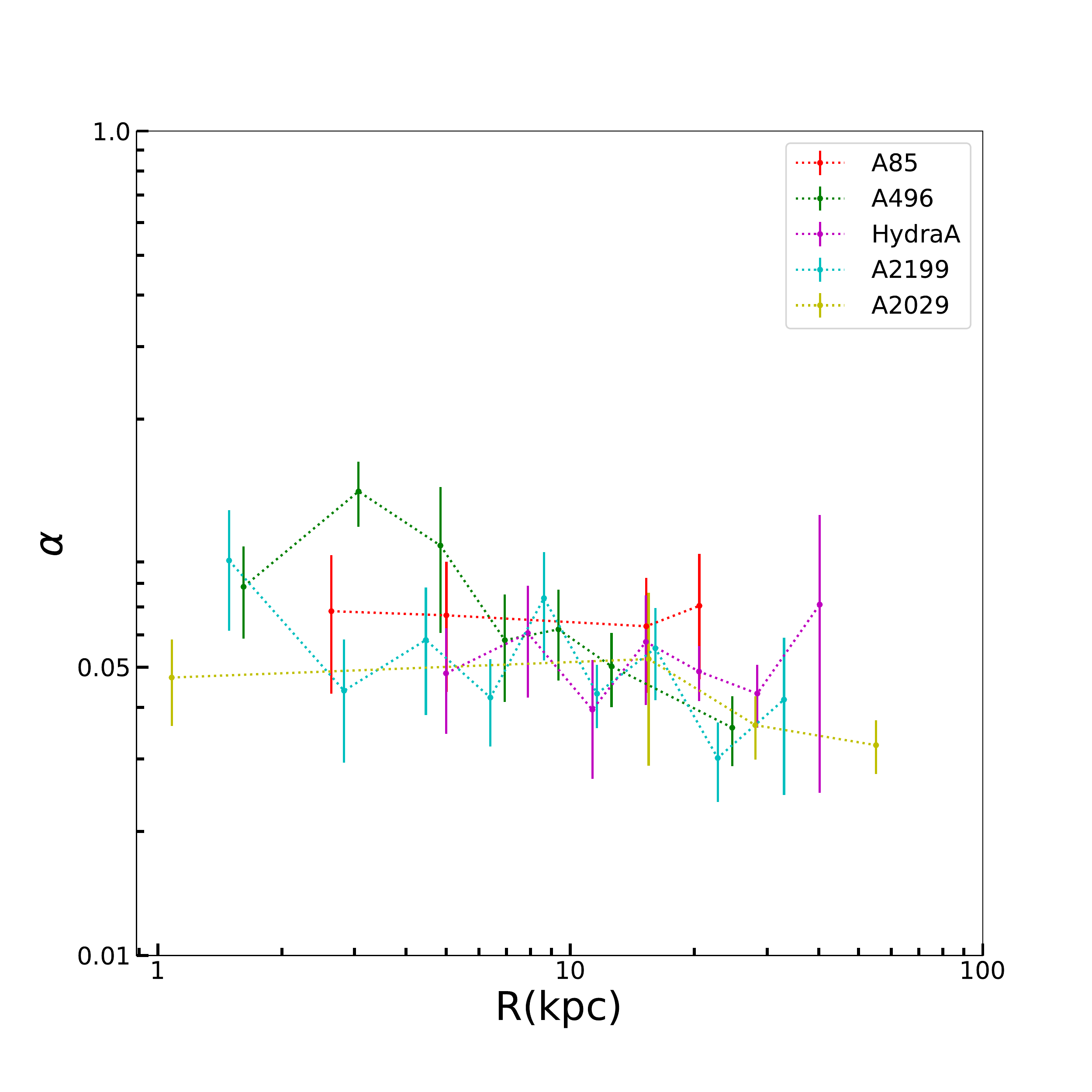}
	\includegraphics[scale = 0.35]{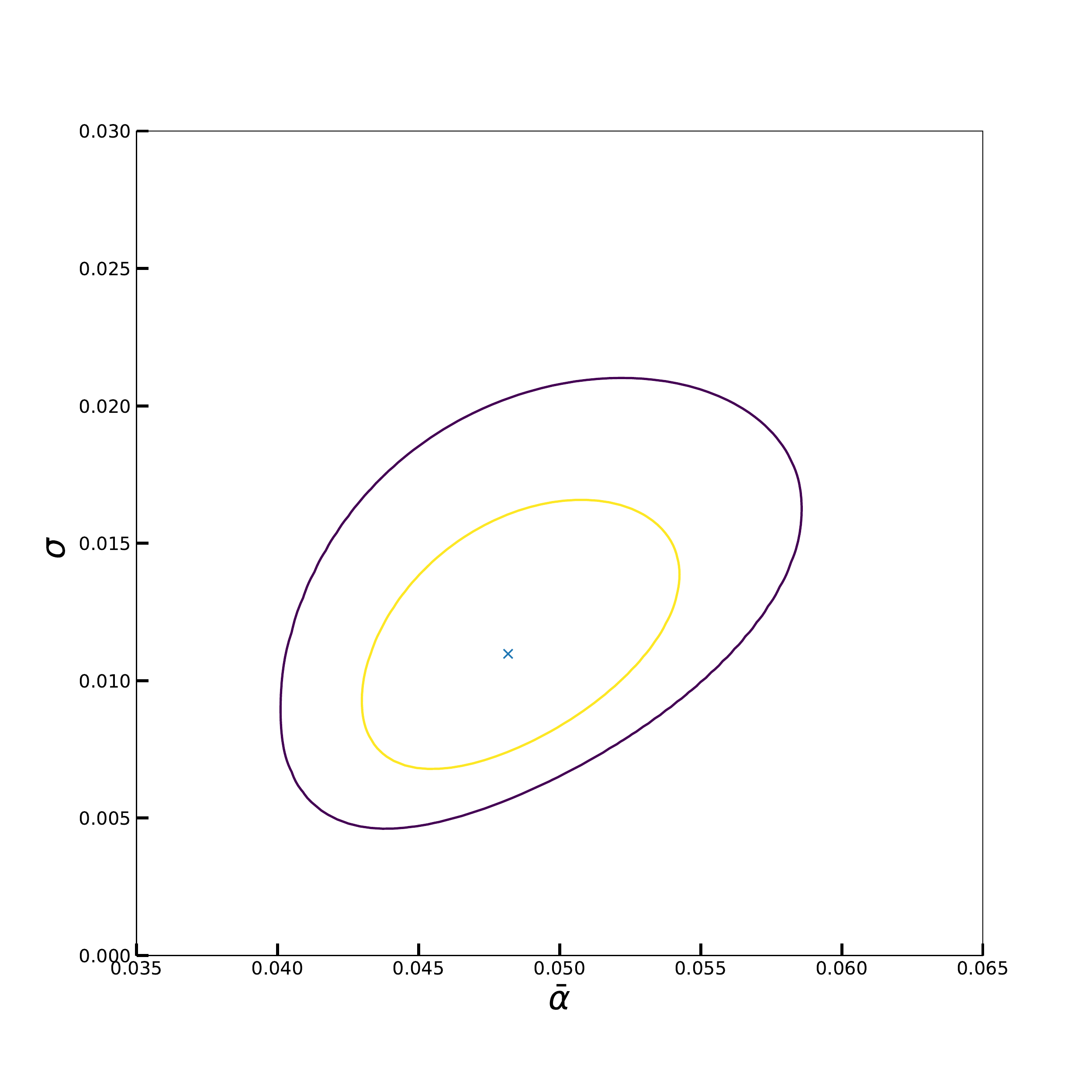}
	\label{zetafig}
	\caption{(a) (top) The measured MHD/turbulent viscous heating parameter $\alpha$, defined as the square of the ratio of sounds crossing to viscous dissipation time $\alpha = (t_{\rm heat}/t_{\rm sound})^{2/3}$. These values are inferred by fitting our CpH model to the spectra and fluxes of deprojected X-ray data from Chandra clusters. (b) (bottom) The 68\% and 95\% confidence regions for the mean $\bar{\alpha}$ and gaussian intrinsic scatter $\sigma$, assuming $\alpha=\bar{\alpha}\pm \sigma$. }
\end{figure}

We can now plug our best-fit CpH spectral model into Eq. (\ref{ALPHA}) to find the MHD/turbulent viscosity parameter $\alpha$, which is plotted in  Figure \ref{zetafig}. We find that a value of $\alpha \simeq 0.05$ (with a small intrinsic scatter of $\pm 10\%$)  in the cluster cores (< 20-30 kpc, where CpH model can give a satisfactory fit to spectrum in Table \ref{Best_fit_tabel}). More precisely, measured $\alpha$'s are consistent with a gaussian distribution with mean $\bar{\alpha}$ and scatter $\sigma$ (see Figure \ref{zetafig}b)

\begin{equation}
	\bar{\alpha} = 0.048 ^{+ 0.005}_{-0.006} ,~~ \sigma = 0.011 ^{+ 0.004}_{-0.006},
	\end{equation}
where errors reflect 1$\sigma$ uncertainties.

These values are consistent with viscosity parameters in shearing box accretion disk simulations \cite{2016MNRAS.457..857S}, as well as simulated \citep[e.g.,][]{2013A&A...559A..78G} or observed turbulent energy fraction in cluster cores \citep{2014Natur.515...85Z,2016MNRAS.458.2902Z,Hitomi2016}.

\begin{figure}
\includegraphics[scale = 0.35]{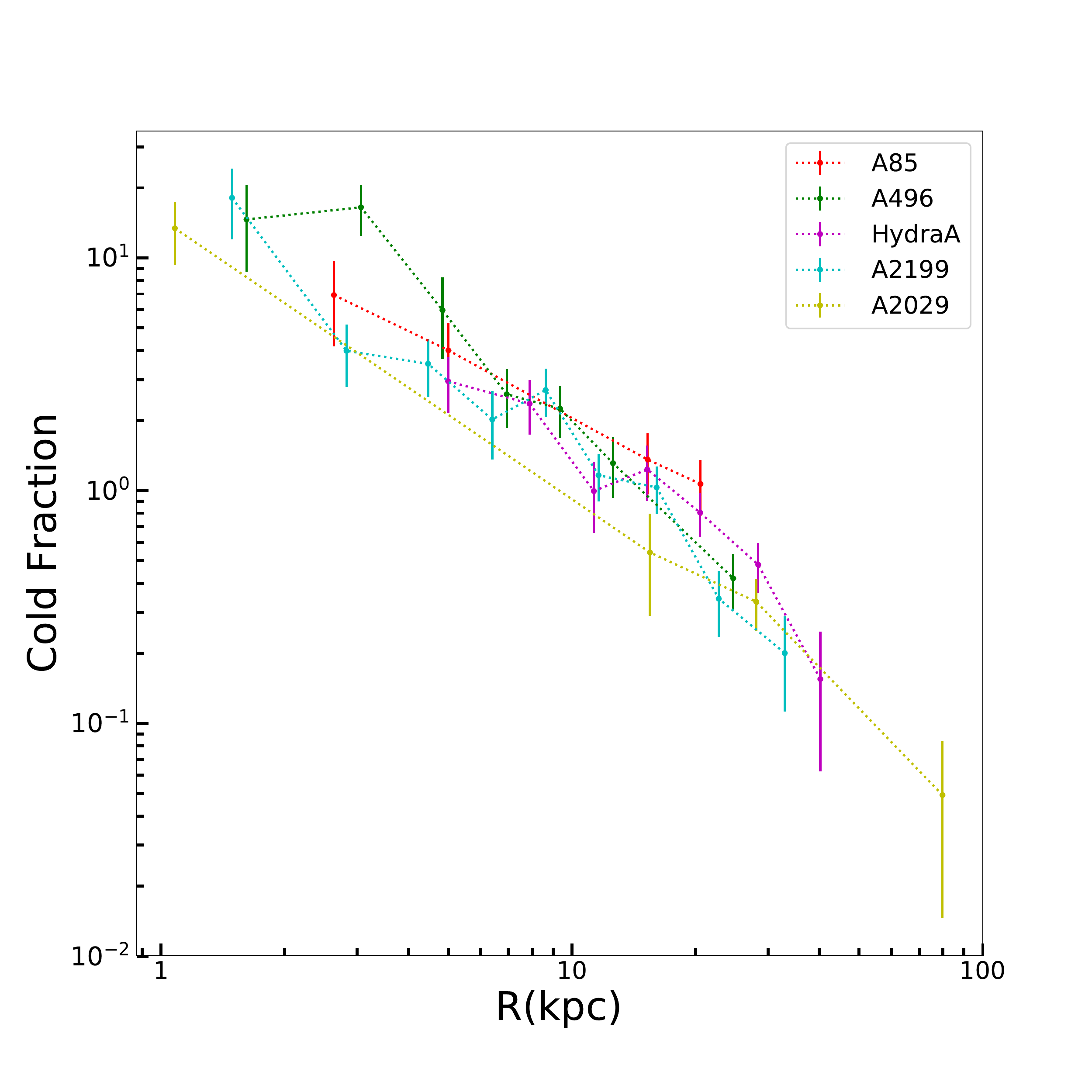}
	\label{cold}
	\caption{ The predicted  cold to hot gas density fraction  accumulated over 7.7 Gyr (i.e. since $z=1$) , assuming our best-fit model and steady state.}
\end{figure}

Considering that our proposed model is successful in representing the observed X-ray spectra, we can compare the mass deposition rate of our CpH model with the observed star formation rate. The key parameter for mass deposition rate is $\dot{M}$, which is directly provided by the spectral fitting in the XSPEC. 
 Another useful quantity to be calculated is the ratio of cold to hot gas density, or cold fraction, which is $7.7 ~{\rm Gyr} \times A$. We estimate the cold fraction over a period of $\sim$ 7.7 Gyr (i.e. since $z=1$) for each cluster. The cold fraction is shown in Figure \ref{cold}, which indicates that most of the cold gas is accumulated within the inner 10 kpc of cluster cores, where its density dominates the hot gas by up to an order of magnitude (we have used only ``good fits'' from Table \ref{Best_fit_tabel}, in Figs \ref{zetafig} and \ref{cold}).

\begin{figure}
	\centering

\includegraphics[scale = 0.35]{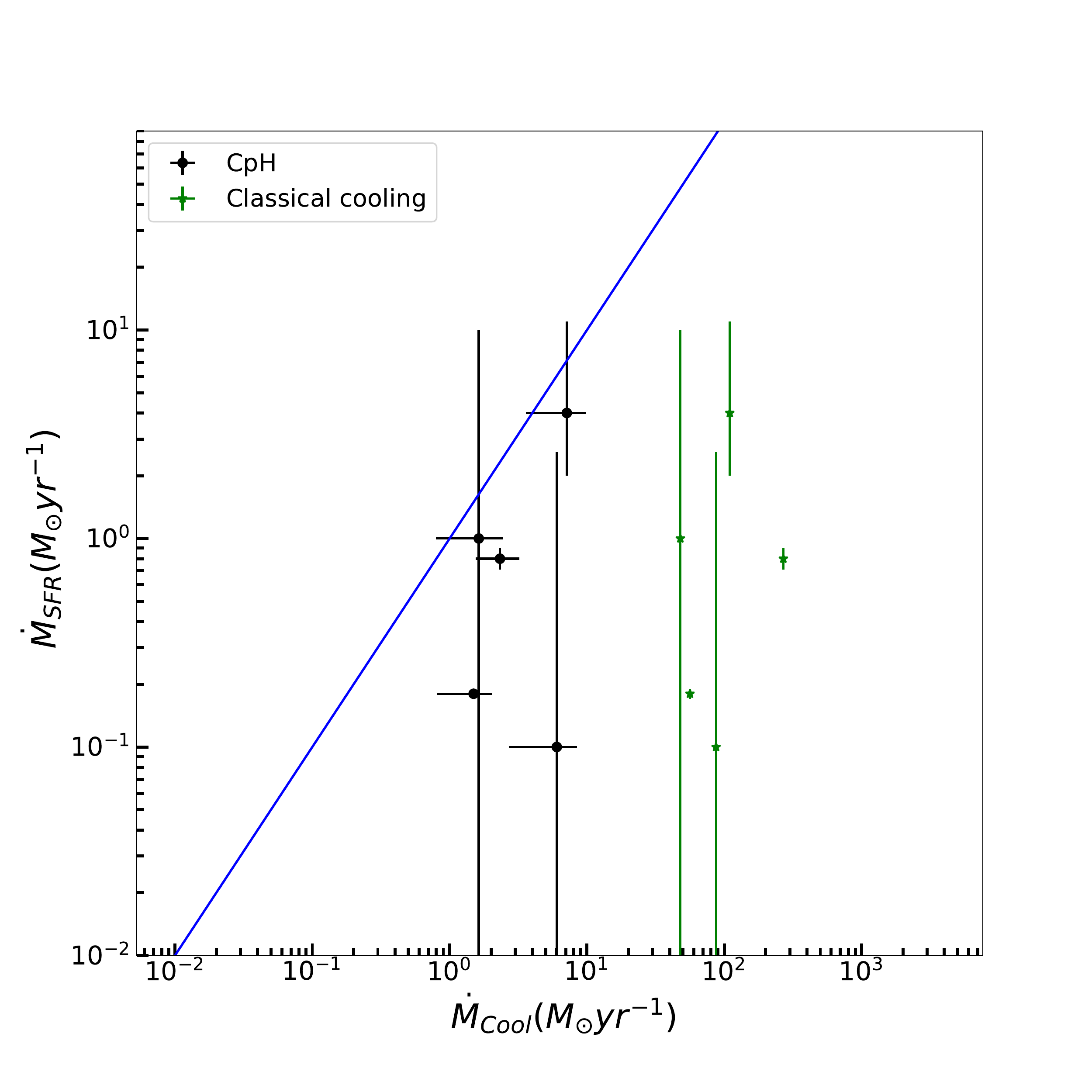}
\caption{The observed star formation rates \citep{McDonald2018} against mass deposition rates. Black points are results of using CpH and green stars are mass deposition rate within the cooling radius of $t_{cool}\leq 3Gyr$ \citep{McDonald2018} for each cluster. The blue line represents  $\dot{M}_{SFR}=\dot{M}_{cool}$.}
	\label{key}
\end{figure}

Let us assume that the cooled gas is used as fuel for star formation. We can define cooling efficiency as $\epsilon_{cool} \equiv \frac{\dot{M}_{SFR}}{\dot{M}_{cool}}$ from which we can specify how well AGN feedback can offset the runaway cooling. We have plotted $\dot{M}_{SFR}$ against $\dot{M}_{cool}$ for each cluster in Figure \ref{key}, showing that $\langle \dot{M}_{cool}\rangle_{\rm CpH} = 3.7^{+0.8}_{-1.0} M_\odot/{\rm yr}$, while $\langle \dot{M}_{cool}\rangle_{\rm classical} \equiv M_{gas}(r<r_{cool})/t_{cool} = 114 \pm 0.5 M_\odot/{\rm yr}$. Here in calculating $\dot{M}_{cool}$ for each cluster in the context of CpH model, we have considered and added the mass deposition rate of annuli that satisfy  $A^{-1} < 3 Gyr $. This condition ensures that the thermodynamic equilibrium in an annulus at temperature $T^*$ has been established. 
On the other hand, in calculating the classical mass deposition rate $\dot{M}_{cool}$,	the  cooling radius $r_{cool}$ is defined as the radius within which cooling time $t_{cool} < 3$ Gyr  \citep{McDonald2018}. The observed star formation rates and classical cooling rates are reported in Table \ref{Observations_Table} in which the mean value of $\dot{M}_{SFR}$ is $1.2^{+2.3}_{-0.5}$.
As a result we can calculate the cooling efficiency for underlying clusters which leads to $\epsilon_{cool} = 0.33^{+0.63}_{-0.15}$ for CpH and $\epsilon_{cool} = 0.01^{+0.02}_{-0.004}$  using results of   \cite{McDonald2018}. Based on these results we infer a much larger efficiency for star formation from cooling in the CpH model, i.e.  $\epsilon_{cool}^{CpH} /\epsilon_{cool}^{classical}  \sim 0.33/0.01 \sim 33 $ .

It worth mentioning that a clear advantage of our model compared to  {\it mkcflow} model is that we have included the contribution of hot gas in all available temperatures and we have not omitted low temperature plasma while fitting the spectrum. In other words, even using all the hot gas to fit the spectrum, we have a consistent predicted viscosity parameter with observation and simulations, which can only be achieved for {\it mkcflow} model by introducing an unphysical low temperature cut-off.

\section{conclusion}
We have shown that by introducing a heating time-scale, modulated by sound-crossing time across the cluster, we can provide superior fits to the X-ray spectra of cores of galaxy clusters using a modified cooling(+heating) flow model, with the same number of parameters as the single-temperature model (CpH model is preferred to the {\it mekal} model at 7.5$\sigma$ level). As a byproduct, we find an MHD/turbulent viscosity parameter of $\alpha =0.048 \pm 0.01$ (intrinsic scatter) $\pm 0.006$ ($1\sigma$ error), which is consistent with simulations and direct observations of turbulent energy fraction in cluster cores. The model can predict the cooling efficiency and the cold fraction by assuming a steady state for $\sim $ 7.7 Gyr (since $z \simeq 1$),  providing a concrete solution to the {\it cooling flow problem}. The cooling efficiency of CpH is $\epsilon_{cool} = 0.33^{+0.63}_{-0.15}$ which is much more larger than classic cooling efficiency $\epsilon_{cool} = 0.01^{+0.02}_{-0.004}$ within cooling radius of $ r(t_{cool}<3Gyr)$.  Furthermore, this model unifies the picture of astrophysical accretion, as a balance of MHD turbulent heating and cooling, across 16 orders of magnitude in scale (from neutron stars and X-ray binaries to cluster cores).

\begin{table*}
	\centering
	
	\begin{tabular}{|c||c|c|c|c|c|c|c|c|c|}
		\hline\hline
		Cluster & $R(kpc)$ &   $\chi^2_{CpH} $   & $\chi^2_{{\it mekal}}$     &  $T^*(keV)$           &$ T_{{\it mekal}}(keV) $ &   $\dot{M}(M_{\odot}/yr) $ & $L_{x}(erg/s)$ & $Net\_counts$ & $N_{d.o.f}$   \\
		
		\hline

		& 1.49$\pm 0.6$ & 62 & 66 & $ 1.4^{+0.4}_{-0.5}$ & 1.3 $^{+0.3}_{-0.4}$& $ 0.1^{+0.04}_{-0.05}$& $6.76^{+0.86}_{-0.88}\times10^{40}$ &3300  	&\\
	& 2.83$\pm0.74$ & 66 & 66 & $ 3.0^{+1.1}_{-0.7}$ &2.5$^{+0.7}_{-0.4}$ & $ 0.12^{+0.06}_{-0.07}$ & $2.89^{+0.19}_{-0.22}\times10^{41}$ &	7203&\\
	& 4.46$\pm0.89$ & 60 & 61 & $ 2.8^{+0.7}_{-0.5}$&2.4$^{+0.5}_{-0.3}$& $ 0.26^{+0.09}_{-0.12}$ & $ 5.35^{+0.34}_{-0.27}\times10^{41}$ &11695	&\\
	& 6.40$\pm1.04$ & 61 & 61 &$ 2.5^{+0.4}_{-0.3}$ &2.3$^{+0.4}_{-0.2}$ &$ 0.14^{+0.16}_{-0.11}$ & $7.7^{+0.36}_{-0.38}\times10^{41}$ &16671	&\\
	A2199	& 8.63 $\pm1.19$& 76 & 82  &$ 2.9^{+0.3}_{-0.5}$ &2.5$^{+0.2}_{-0.2}$ &$ 0.75^{+0.19}_{-0.28}$ & $16.5^{+0.4}_{-0.5}\times10^{41}$ & 21142	& 64\\
	& 11.61 $\pm1.79$& 62 & 62 &$ 2.9^{+0.3}_{-0.3}$ &2.7$^{+0.2}_{-0.2}$ & $ 0.27^{+0.28}_{-0.20}$& $21.0^{+0.4}_{-0.5}\times10^{41}$ & 28883	&\\
	& 16.07$\pm2.68$ & 63 & 66 &$ 2.8^{+0.5}_{-0.6}$ &2.5$^{+0.3}_{-0.3}$ & $ 0.73^{+0.33}_{-0.41}$ & $18.7^{+0.65}_{-0.52}\times10^{41}$ &38885	&\\
	& 22.77$\pm4.01$ & 57 & 56 &$ 3.3^{+0.6}_{-0.4}$ &3.1$^{+0.3}_{-0.3}$ &$ 0.33^{+0.33}_{-0.25}$& $32.5^{+0.74}_{-0.67}\times10^{41}$ & 68880	&\\
	& 32.97 $\pm6.18$& 47 & 48 &$ 4.4^{+0.5}_{-0.5}$ &3.9$^{+0.3}_{-0.3}$ & $ 1.1^{+0.7}_{-0.7}$& $76.5^{+0.89}_{-0.93}\times10^{41}$ & 127922	&\\
	& 48.2$\pm9.08$ & \textit{\textcolor{red}{101}} &  \textit{\textcolor{red}{105}} &$ 4.7^{+0.2}_{-0.3}$ &4.4$^{+0.1}_{-0.1}$ &$ 2.2^{+1.1}_{-1.1}$ & $243^{+1.4}_{-1.3}\times10^{41}$ & 216253	&\\

		\hline
		
			& 5.00$\pm 1.3$ & 19 & 28 & $3.7^{+0.9}_{-0.7}$&$3.1^{+0.5}_{-0.4}$ &$1.6^{+0.3}_{-0.4}$ & $9.5^{+0.35}_{-0.33}\times10^{41} $ &7241 & \\
		& 7.89$\pm1.6 $ & 17 & 20 &$4^{+1.0}_{-0.7}$ & $3.3^{+0.5}_{-0.4}$ & $1.5^{+0.4}_{-0.6}$ &$16.2^{+0.64}_{-0.5}\times10^{41} $ &9862	&\\
		& 11.31 $\pm1.8 $ & 21 & 22 &$3.0^{+0.7}_{-0.6}$ &$2.5^{+0.4}_{-0.4}$ &$1.5^{+0.6}_{-0.9}$ &$16.0^{+0.5}_{-0.6}\times10^{41} $ & 16102	&\\
		& 15.25$\pm 2.1$ & 27 & 27 & $5^{+2.0}_{-1.4}$ &$3.7^{+1.0}_{-0.8}$ &$1.1^{+0.4}_{-0.6}$ & $20.6^{+0.9}_{-1.9}\times10^{41} $ & 20226&\\
		HydraA	&  20.51$\pm 3.2$ & 14 & 14 &$3.2^{+0.7}_{-0.5}$ &$2.8^{+0.4}_{-0.3}$ &$2.5^{+1.4}_{-1.7}$ & $32.1^{+1.0}_{-1.0}\times10^{41} $  &26833 & 21	\\
		&28.40 $\pm4.7 $ & 22 & 23 & $3.5^{+0.8}_{-0.3}$ &$3.1^{+0.3}_{-0.2}$ & $3.7^{+2.6}_{-2.6}$ & $73.1^{+1.2}_{-1.1}\times10^{41} $ & 44738	&\\
		& 40.24$\pm 7.1$ & 25 & \textit{\textcolor{red}{33}} &$4.4^{+0.6}_{-0.5}$ &$3.7^{+0.3}_{-0.2}$ &$7.7^{+3.3}_{-3.6}$ & $124^{+1.6}_{-1.7}\times10^{41} $ &68412 	&\\
		&   58.25 $\pm10.9 $ &\textit{\textcolor{red}{36}}  & \textit{\textcolor{red}{42}} & $4.5^{+0.4}_{-0.1}$ &$4.2^{+0.1}_{-0.1}$ & $8.2^{+6.1}_{-3.9}$ & $258^{+1.9}_{-2.1}\times10^{41} $ & 111067	&\\
		& 85.20$\pm 16.0$ & \textit{\textcolor{red}{67}} & \textit{\textcolor{red}{87}} & $ 4.6^{+0.3}_{-0.2}$ &$4.3^{+0.1}_{-0.1}$ &  $13.4^{+6.7}_{-5.6}$  & $409^{+2.7}_{-2.0}\times10^{41} $ & 145225	&\\
		& 125.2$\pm 23.9$ &\textit{\textcolor{red}{45}} & \textit{\textcolor{red}{64}} &$ 4.6^{+0.3}_{-0.2}$ &$4.2^{+0.1}_{-0.1}$ &$ 14.4^{+7.2}_{-7.2}$ & $388^{+2.0}_{-1.9}\times10^{41} $ &153239 	&\\

		\hline
		& 1.08$\pm1.0$ & 63 & 63 &$2.6^{+0.3}_{-0.65}$ &2.2$^{+0.4}_{-0.3}$ &$0.21^{+0.24}_{-0.17}$ & $4.2^{+0.1}_{-0.18}\times10^{41}$ & 1993	&\\
		& 3.60$\pm1.44$ & 103&103 &$3.0^{+0.6}_{-0.4}$&2.8$^{+0.4}_{-0.3}$ &$0.31^{+0.40}_{-0.24}$& $35.2^{+2.0}_{-3.6}\times10^{41}$ & 12208	&\\
		& 6.84$\pm1.8$ &\textit{\textcolor{red}{104}} & \textit{\textcolor{red}{104}}& $3.1^{+0.5}_{-0.2}$&2.9$^{+0.5}_{-0.3}$ &$0.50^{+0.67}_{-0.40}$& $49.7^{+1.1}_{-1.7}\times10^{41}$ & 13500	&\\
		& 10.80$\pm2.16$ &\textit{\textcolor{red}{105}} &\textit{\textcolor{red}{106}} & $11.7^{+2.7}_{-3.4}$&7.8$^{+3.2}_{-1.8}$ &$1.3^{+0.4}_{-0.6}$ & $91.4^{+2.8}_{-3.0}\times10^{41}$ & 22702	&\\
		A2029	& 15.49 $\pm2.5$& 98 & 97 & $5.6^{+2.2}_{-1.2}$ &4.5$^{+0.9}_{-0.6}$ &$1.8^{+1.6}_{-1.4}$ & $12.9^{+0.4}_{-0.3}\times10^{42}$ &  28949	&  86 \\
		& 20.89$\pm2.88$ & \textit{\textcolor{red}{110}} &\textit{\textcolor{red}{110}} &$9.6^{+4.1}_{-3.0}$ &6.4 $^{+1.2}_{-1.0}$& $2.3^{+1.3}_{-1.6}$ & $19.1^{+0.3}_{-0.6}\times10^{42}$ & 34104	&\\
		& 28.09$\pm4.32$ & 77 &77 & $7.2^{+2.4}_{-1.3}$ & 6.0$^{+0.8}_{-0.5}$ &$2.9^{+3.0}_{-2.2}$ & $33.3^{+0.5}_{-0.5}\times10^{42}$ & 56203	&\\
		& 38.90 $\pm6.48$&\textit{\textcolor{red}{117}} &\textit{\textcolor{red}{117}} &$8.1^{+1.7}_{-0.8}$ &6.9$^{+0.5}_{-0.5}$ & $4.0^{+3.7}_{-2.9}$ & $63.8^{+0.6}_{-0.5}\times10^{42}$ & 88105	&\\
		& 55.10$\pm9.72$ & 101 & 101 &$ 7.2^{+1.8}_{-1.9}$ &7.4 $^{+0.5}_{-0.4}$& $ 1.7^{+1.7}_{-1.3}$ & $99.8^{+0.8}_{-0.3}\times10^{42}$ & 128336	& \\
		& 79.77$\pm14.95$ & \textit{\textcolor{red}{104}} & 102 &$8.9^{+1.0}_{-0.8}$ &7.8$^{+0.4}_{-0.3}$ &$3.6^{+6.4}_{-3.5}$ & $165^{+1.0}_{-1.0}\times10^{42}$ & 171978	&\\
		
		\hline
		&1.61$\pm0.65$ & 5 & 9 & $1.7^{+0.6}_{-0.4}$ & 1.4$^{+0.44}_{-0.19}$ & $0.15^{+0.06}_{-0.06}$ & $1.3^{+0.1}_{-0.1}\times10^{41}$ & 533	&\\
		&3.07 $\pm0.81$ & 4 & 10 & $1.3^{+0.2}_{-0.2}$ &1.1$^{+0.2}_{-0.1}$ & $0.47^{+0.10}_{-0.15}$ & $2.7^{+0.2}_{-0.2}\times10^{41}$ & 2156	&\\
		&4.84$\pm0.97$  & 6 & 8 & $1.8^{+0.5}_{-0.2}$ &1.7$^{+0.2}_{-0.2}$ &  $0.50^{+0.17}_{-0.22}$ & $5.6^{+0.4}_{-0.4}\times10^{40}$ & 4680	&\\
		& 6.94$\pm1.13$ & 7 & 7 & $2.9^{+1.2}_{-0.8}$ &2.3$^{+0.6}_{-0.3}$ & $0.38^{+0.20}_{-0.25}$ & $10.1^{+0.6}_{-0.5}\times10^{41}$ & 7361	&\\
		A496	& 9.36$\pm1.29$& 6 & 6 & $2.8^{+0.8}_{-0.6}$ &2.3$^{+0.5}_{-0.2}$ & $0.51^{+0.37}_{-0.33}$ & $15.9^{+0.5}_{-0.6}\times10^{41}$ & 10129	& 10\\
		& 12.59$\pm1.94$ & 12 & 11 & $2.4^{+0.8}_{-0.2}$ &2.2$^{+0.2}_{-0.1}$ & $0.52^{+0.37}_{-0.39}$ & $32.6^{+0.8}_{-0.7}\times10^{41}$ & 13736	&\\
		& 17.43$\pm2.90$ & \textit{\textcolor{red}{24}} & \textit{\textcolor{red}{24}} & $2.9^{+0.3}_{-0.3}$ &2.7 $^{+0.1}_{-0.1}$& $0.37^{+0.26}_{-0.29}$& $48.6^{+1.0}_{-0.8}\times10^{41}$ & 21962	&\\
		& 24.69$\pm4.36$ & 7 & 6 & $3.2^{+0.4}_{-0.3}$ &3.0$^{+0.3}_{-0.2}$ & $0.55^{+0.55}_{-0.44}$ & $80.4^{+1.2}_{-1.1}\times10^{41}$ & 31017	&\\
		& 35.74$\pm6.70$ & \textit{\textcolor{red}{24}} & \textit{\textcolor{red}{23}} & $4.1^{+0.3}_{-0.5}$ &3.8$^{+0.2}_{-0.2}$ & $0.71^{+0.71}_{-0.57}$ & $122^{+1.2}_{-1.3}\times10^{41}$ & 46112	&\\
		& 52.29$\pm9.84$ & \textit{\textcolor{red}{19}} & \textit{\textcolor{red}{19}} & $4.3^{+0.6}_{-0.3}$ &  4.0$^{+0.2}_{-0.2}$ & $1.1^{+1.1}_{-1.0}$ & $194^{+1.7}_{-1.1}\times10^{41}$ & 64190	&\\
		
		\hline
		
		& 2.63 $\pm1.1$& 23 & 22  & $2.4^{+1.1}_{-0.6}$ &1.98$^{+0.7}_{-0.3}$ & $0.34^{+0.18}_{-0.17}$ & $5.2^{+0.4}_{-0.4}\times10^{41}$ & 3503	&\\
		& 5.0$\pm1.32$  & 23 & 22 & $3^{+1.1}_{-0.7}$ &2.4$^{+0.7}_{-0.4}$ & $0.68^{+0.29}_{-0.36}$ & $14.7^{+0.8}_{-0.9}\times10^{41}$ & 8458	&\\
		&7.90$\pm1.58$& \textit{\textcolor{red}{38}} & \textit{\textcolor{red}{39}} & $2.7^{+0.4}_{-0.4}$ &2.2$^{+0.4}_{-0.2}$ & $1.69^{+0.51}_{-0.79}$ & $32.3^{+1.2}_{-1.4}\times10^{41}$ & 11147	&\\
		& 11.33$\pm1.84$ &\textit{\textcolor{red}{98}} &\textit{\textcolor{red}{98}} & $2.6^{+0.4}_{-0.3}$ &2.3$^{+0.1}_{-0.1}$ & $1.96^{+0.90}_{-1.2}$ & $42.5^{+1.3}_{-1.3}\times10^{41}$ & 14413	&\\
		A85 & 15.28$\pm2.11$ & 32  &  32 & $3.6^{+1.1}_{-0.8}$ &2.8$^{+0.6}_{-0.4}$ & $1.4^{+0.6}_{-0.8}$ & $39.4^{+1.2}_{-2.2}\times10^{41}$ & 16205	& 24\\
		& 20.55$\pm3.16$ & 22 & 28 & $3.3^{+0.9}_{-0.6}$ &2.8$^{+0.6}_{-0.4}$ & $2.3^{+0.6}_{-0.6}$ & $52.0^{+2.2}_{-1.5}\times10^{41}$ & 26093	&\\
		& 28.45$\pm4.74$ &\textit{\textcolor{red}{93}} &\textit{\textcolor{red}{93}}& $4.4^{+1}_{-0.8}$ &3.8$^{+0.3}_{-0.3}$ & $3.7^{+0.7}_{-2.9}$ & $143^{+2.0}_{-2.4}\times10^{41}$ & 46348	&\\
		& 40.31$\pm7.11$&\textit{\textcolor{red}{73}} &\textit{\textcolor{red}{ 73}} & $5.8^{+1.3}_{-1}$ &4.6$^{+0.2}_{-0.2}$ & $3.3^{+2.2}_{-2.2}$ & $223^{+3}_{-2}\times10^{41}$ & 68838	&\\
		&  58.36 $\pm10.93$&\textit{\textcolor{red}{198}} &\textit{\textcolor{red}{198}} & $6.4^{+1.2}_{-1}$ &5.1$^{+0.2}_{-0.2}$ & $5.3^{+3.2}_{-4.9}$ & $381^{+3.2}_{-2.2}\times10^{41}$ & 105746	&\\
		& 85.36$\pm16.10$ &\textit{\textcolor{red}{278}} & \textit{\textcolor{red}{278}}& $6.7^{+1.3}_{-0.7}$ &5.8$^{+0.2}_{-0.2}$ & $3.4^{+4.5}_{-2.2}$ & $556^{+4.7}_{-3.5}\times10^{41}$ & 134958	&\\
		\hline

	\end{tabular}
	\caption{Calculated best-fit $\chi^2$ of our CpH model and {\it mekal} single-temperature model,  with the best-fit parameters of fitting for the first ten annuli. $T^*$ and $\dot{M}$ are best fit parameters of our CpH model.
		The red (italic) fonts suggest that the fit is outside the 90\% expected range for the (reduced) $\chi^2$ for  $N_{d.o.f}$, i.e. it is not a good fit. We see that our CpH model typically provides a lower $\chi^2$, or a better fit, in cluster cores. }
	\label{Best_fit_tabel}
\end{table*}

\acknowledgements

We would like to thank Shane Davis, Mike Hogan, Brian McNamara, Faerlin Pulido, Adrian Vantyghem, and ... for useful discussions. MZH thanks participants in the weekly cosmology group meeting at Perimeter Institute for their useful comments. We also would like to thank Mike Hogan for providing us with the data used in \cite{Hogan17} and  we thank the anonymous referee for his/her review and useful comments. The scientific results reported in this article are based on observations made by the Chandra X-ray Observatory and has made use of software provided by the Chandra X-ray Center (CXC) in the application packages CIAO, ChIPS, and Sherpa.
This research is supported in part by the University of Waterloo, National Science and Engineering Council of Canada (NSERC), and Perimeter Institute for Theoretical Physics. Research at Perimeter Institute is supported by the Government of Canada through the Department of Innovation, Science and Economic Development Canada and by the Province of Ontario through the Ministry of Research, Innovation and Science.

\appendix

\begin{figure}

	\includegraphics[scale = 0.35]{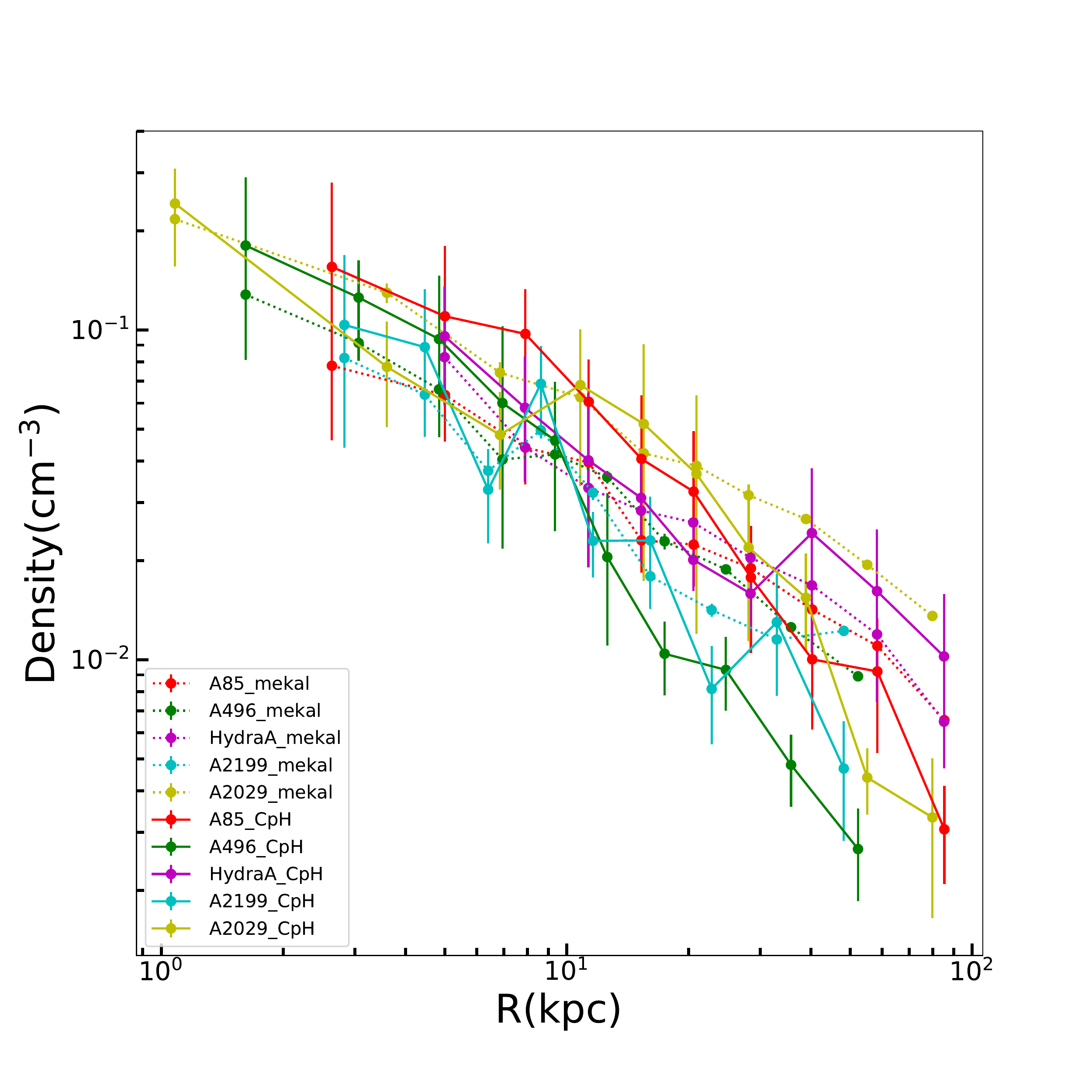}
	\includegraphics[scale = 0.35]{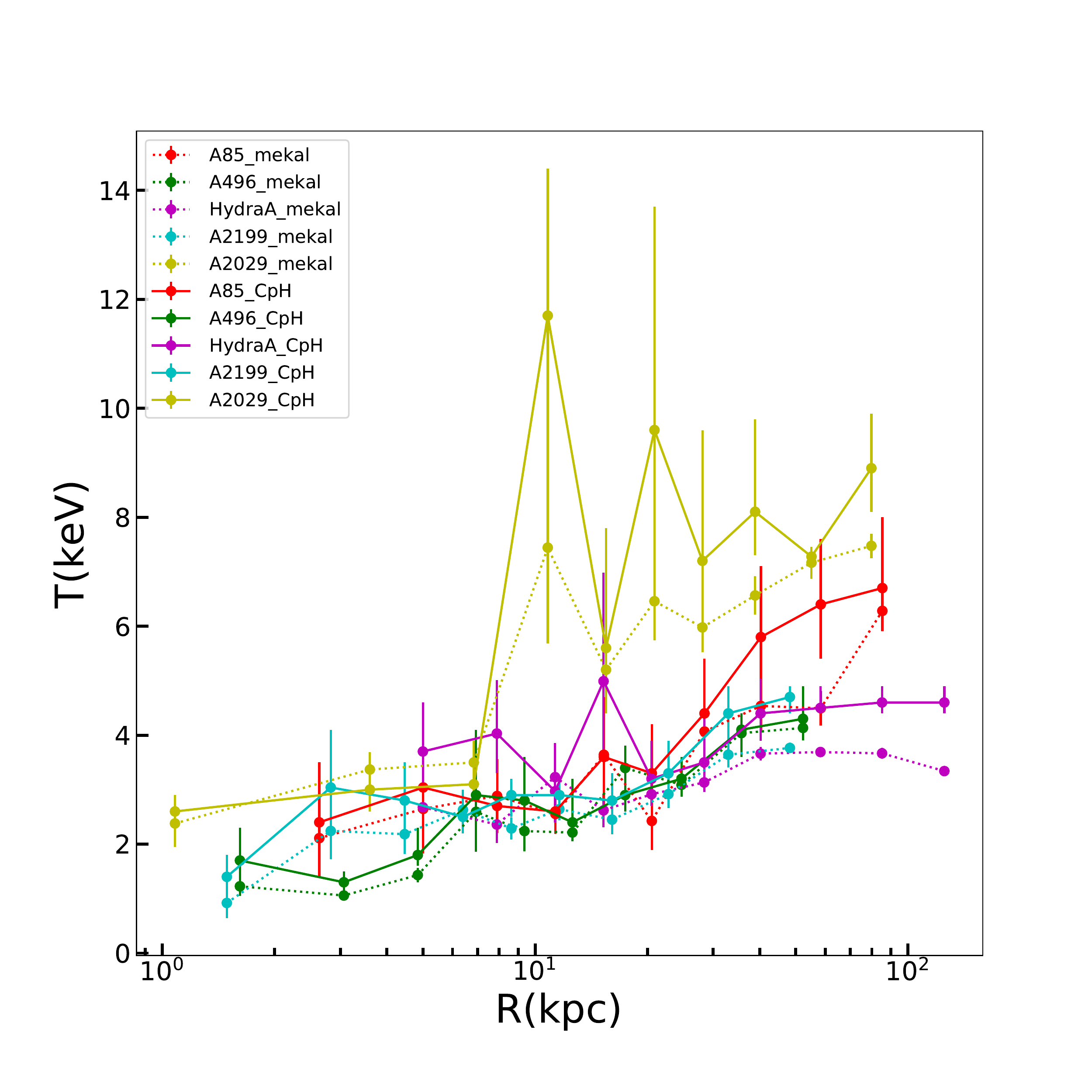}
	\caption{ Deprojected density and temperature profiles of CpH (solid lines) compared with what \cite{Hogan17} found using Mekal (dotted lines).}
	\label{profile}
\end{figure}

\begin{figure}
	\centering
	\includegraphics[scale = 0.35]{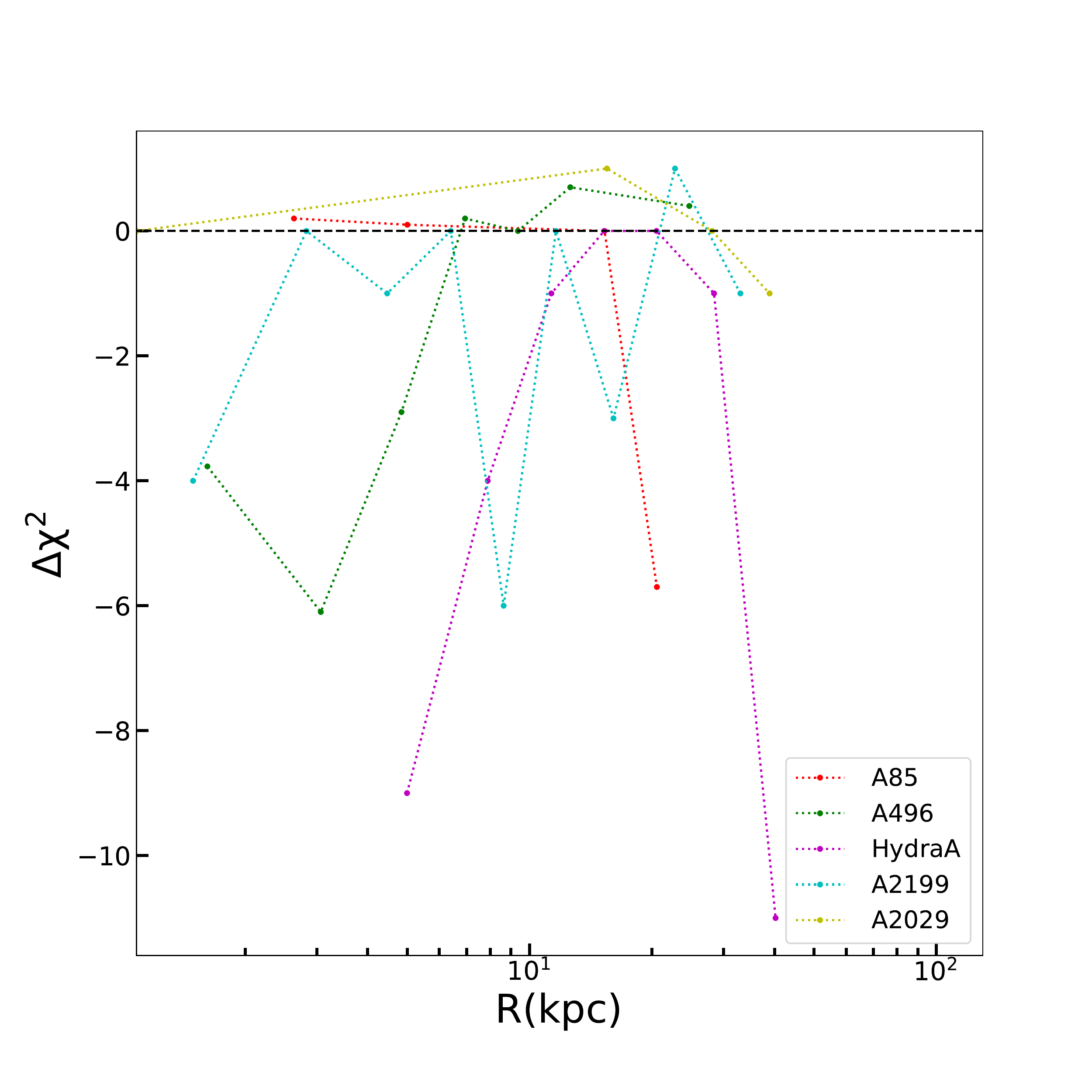}
	\caption{$\Delta \chi^2 $ profile based on  good fits in Table \ref{Best_fit_tabel}. It is apparent from this plot that the majority of bins have lower $\chi^2$ for CpH with respect to Mekal.}
	\label{chi2plot}
\end{figure}

In the appendix we are providing temperature, density and $
\Delta \chi^2$ profiles of CpH and Mekal model. Figure \ref{profile} shows the close proximity of the density profiles in CpH and Mekal Model (Data of Mekal profiles are extracted from \cite{Hogan17}). As can be inferred from Figure \ref{profile} the peak temperature profiles of CpH  and Mekal temperatures profiles show almost the same trend. 
In order to verify that the CpH model provides an improved fit over the Mekal model, we present the $\Delta\chi^2 = \chi^2_{CpH} -  \chi^2_{Mekal}$  profile in Figure \ref{chi2plot} in which we are using only good fits (black bins in Table \ref{Best_fit_tabel}). 

\end{document}